\documentclass[12pt,twoside,a4paper]{article}
\usepackage{amsmath,amssymb,latexsym,theorem,bbm,natbib,epsfig,color,
  footnote} 
\def\crps{\mathop{\hbox{\rm CRPS}}}
\def\logs{\mathop{\hbox{\rm LogS}}}
\def\twcrps{\mathop{\hbox{\rm twCRPS}}}
\def\twcrpss{\mathop{\hbox{\rm twCRPSS}}}

\numberwithin{equation}{section}

\title{Mixture EMOS model for calibrating ensemble forecasts of wind speed}

\author{S\'andor Baran$^{\mathrm{a}}$ and 
    Sebastian Lerch$^{\mathrm{b,c}}$ \\ [2mm]
{\small
$^{\mathrm a}$Faculty of Informatics, University of Debrecen, Hungary} \\
{\small $^{\mathrm b}$Heidelberg Institute for Theoretical Studies, Germany} \\
{\small $^{\mathrm c}$ {Insitute of Stochastics, Karlsruhe Institute of Technology, Germany}} 
}

\date{}

\begin{document}
\pagestyle{myheadings}

\maketitle

\begin{abstract}
Ensemble model output statistics (EMOS) is a statistical tool for post-processing forecast ensembles of weather variables obtained from multiple runs of numerical weather prediction models in order to produce calibrated predictive probability density functions (PDFs). The EMOS predictive PDF is given by a parametric distribution with parameters depending on the ensemble forecasts. 
We propose an EMOS model for calibrating wind speed forecasts based on weighted mixtures of truncated normal (TN) and log-normal (LN) distributions where model parameters and component weights are estimated by optimizing the values of proper scoring rules over a rolling training period. 
The new model is tested on wind speed forecasts of the 50 member European Centre for Medium-Range Weather Forecasts ensemble, the 11 member Aire Limit\'ee
Adaptation dynamique D\'eveloppement International-Hungary Ensemble Prediction
System ensemble of the Hungarian Meteorological Service and the eight-member University of Washington mesoscale ensemble, 
and its predictive performance is compared to that of various benchmark EMOS models based on single parametric families and combinations thereof.
The results indicate improved calibration of probabilistic and accuracy of point forecasts in comparison with the raw
ensemble and climatological forecasts. The mixture EMOS model significantly outperforms the TN and LN EMOS methods, moreover, it 
provides better calibrated forecasts than the TN-LN combination model and offers an increased flexibility while avoiding covariate selection problems.

\end{abstract}

\markboth{S. Baran and S. Lerch}{Mixture EMOS model for calibrating ensemble forecasts of wind speed} 
\section{Introduction}
  \label{sec:sec1}

In our industrialized world several important applications require reliable and accurate wind speed forecasts. These include, but are not limited to agriculture, aviation or wind energy production. In particular, high wind speeds can cause severe damages to infrastructure and their predictions are important parts of weather warnings. Wind speed forecasts are standard outputs of numerical weather prediction (NWP) models. NWP has traditionally been viewed as a deterministic problem, but over the last decades, a change towards probabilistic forecasts, i.e., forecasts in the form of a full predictive distribution, can be observed. Probabilistic forecasts are important in the context of weather forecasting as they allow for a quantification of the associated uncertainty of the prediction, and further allow for optimal point forecasting by using certain functionals of the predictive distribution \citep[see, e.g., ][]{gneiting11}.

Nowadays, weather services typically produce ensemble forecasts which consist of multiple runs of NWP models that differ in initial conditions and/or the numerical representation of the atmosphere \citep{gr05}. While the transition to ensemble forecasts is an important step towards probabilistic forecasting, ensembles are finite and do not provide full predictive densities. Further, ensemble forecasts are typically underdispersive and subject to systematic bias, they thus require some form of statistical post-processing \citep{gr05,gbr}.   
  
State of the art techniques for statistical post-process\-ing of ensemble forecasts include Bayesian model averaging (BMA) developed by \cite{rgbp}, and ensemble model output statistics (EMOS) or non-homogeneous regression by \cite{grwg}.  The BMA approach uses weighted mixtures of parametric probability density functions (PDFs) which depend on the ensemble forecasts, with the mixture weights being determined based on the performance of the ensemble members in the training period. Possible component choices for wind speed are given by PDFs of Gamma distributions \citep{sgr10} or truncated normal (TN) distributions \citep{bar}. By contrast, the predictive distribution of the EMOS app\-roach is given by a single parametric distribution with parameters depending on the ensemble forecasts. \citet{tg} propose the use of truncated normal distributions for EMOS models of wind speed. Alternative choices are given by generalized extreme value distributions \citep{lt}, log-nor\-mal (LN) distributions \citep{bl}, and 
combinations thereof.
  
The article at hand builds on the EMOS framework and proposes models based on weighted mixtures of TN and LN distributions.   
This new approach allows for combining the advantages of lighter and heavier-tailed distributions, but avoids problems encountered in using previously proposed combination models. Apart from the flexibility, the mixture models further exhibit desirable properties from a theoretical perspective and provide well calibrated and skillful probabilistic forecasts.

The novel EMOS mixture approach is applied to forecasts of maximal wind speed of the 50 member ensemble of the European Centre for Medium-Range Weather Forecasts \citep[ECMWF;][]{ecmwf} and the eight-member University of Washington mesoscale ensemble \citep[UWME; ][]{em05}, and to instantaneous wind-speed forecasts of the 11 member Limited Area Model Ensemble Prediction System of the Hungarian Meteorological Service (HMS) called Aire Limit\'ee Adaptation dynamique D\'eveloppement International-Hungary Ensemble Prediction System \citep[ALADIN-HUNEPS;][]{hagel,hkkr}. The three ensemble prediction system differ in the generation of their members, which is accounted for in the model formulation. The TN model of \citet{tg}, and the LN and TN-LN 
combination models of \citet{bl} serve as benchmark models for the three case studies.

The remainder of this article is organized as follows. Section \ref{sec:sec2} contains a description of the ensembles and observational data. In Section \ref{sec:sec3} the EMOS technique is reviewed and the novel TN-LN mixture models are introduced. Section \ref{sec:sec4} summarizes the results of the three case studies. The article concludes with a discussion in Section \ref{sec:sec5}.

\section{Data}
  \label{sec:sec2}
 
We consider three distinct data sets of ensemble forecasts and corresponding observations which differ both in the stochastic properties of the ensemble as well as the observed wind quantities. The data sets coincide with those used in \citet{bl}. We thus limit our discussion here to a succinct summary of the data and refer to \citet{bl} for a more detailed description.

\subsection{ECMWF ensemble}
  \label{subs:subs2.1}

The ECMWF ensemble consists of 50 exchangeable ensemble members of one day ahead forecasts of 10 m daily maximum wind speed (given in m\,s$^{-1}$) along with corresponding validating observations from 228 synoptic observation stations over Germany. The observations are daily maxima of hourly observations of 10-minute average wind speed measured over the 10 minutes before the hour, where the maxima are taken over the 24 hours corresponding to the time frame of the ensemble forecast. 
The results presented in Section \ref{subs:subs4.1} are based on a verification period
from 1 May 2010 to 30 April 2011, consisting of 83\,220 individual
forecast cases. 

\begin{figure}[t]
\begin{center}
\leavevmode
\epsfig{file=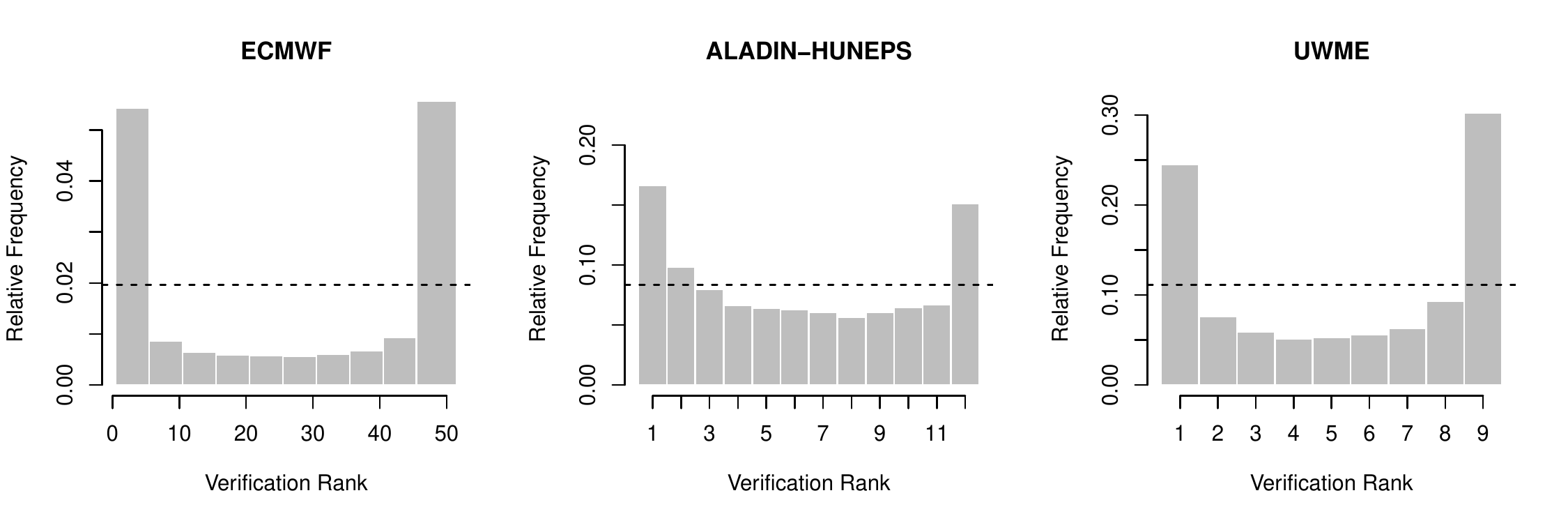,width=16.5cm, angle=0}
\centerline{\hbox to 11 cm{\scriptsize {a) \hfill b) \hfill c)}}}
\end{center}
\caption{Verification rank histograms. a) ECMWF ensemble for the period
  1 May 2010 -- 30 April 2011; b) ALADIN-HUNEPS
  ensemble for the period 1 April 2012 -- 31 March 2013; c) UWME
  for the calendar year 2008.} 
\label{fig:fig1}
\end{figure}

Figure \ref{fig:fig1}a shows the verification rank histogram of the raw
ensemble, that is the histogram of ranks of validating
observations with respect to the corresponding ensemble
forecasts computed from the ranks at all locations and dates
considered \citep[see, e.g.,][Section 7.7.2]{wilks}. The strongly U-shaped histogram indicates a highly underdispersive character of the ECMWF ensemble. The range of the ECMWF ensemble contains the validating observation only in $43.40\,\%$ of all cases
(the nominal value of this coverage is $49/51$, that is 
$96.08\,\%$), verifying the need of statistical post-processing.

\subsection{ALADIN-HUNEPS ensemble}
  \label{subs:subs2.2}

The ALADIN-HUNEPS system of the HMS \citep{hkkr,dljbac} consists of 11 members, 10 exchangeable forecasts initialized from perturbed initial
conditions and one control member from the unperturbed analysis. The data base contains ensembles of 42-h forecasts for 10
m wind speed (given in m\,s$^{-1}$) for 10 major cities in 
Hungary, together with the corresponding validating observations for the one-year period between 1 April 2012 and 31 March 2013.

The validating wind speed measurements are considered as instantaneous values (valid at a given time), however,
they are in fact mean values over the preceding 10 minutes. The model wind speed values are also considered as instantaneous, but they are representatives for a given model time step, which is 5 min in our case. 

Similar to the ECMWF ensemble, the verification rank histogram of
the raw ALADIN-HUNEPS ensemble is far
from the desired uniform distribution (see Figure \ref{fig:fig1}b),
however, it shows a much less 
underdispersive character. The better fit of the ensemble can also be
observed on its coverage value of
$61.21\,\%$ which should be compared with the nominal
coverage of $83.33\,\%$ ($10/12$).

\subsection{University of Washington Mesoscale Ensemble}
  \label{subs:subs2.3}

The UWME covering the Pacific Northwest region of western
North America has eight members that are obtained from different runs of the
fifth generation Pennsylvania State  
University--National Center for Atmospheric Research mesoscale model 
(PSU-NCAR MM5) \citep{grell}. 
Our data base contains ensembles of 48 h forecasts 
and corresponding validating observations of 10 m maximal wind
speed (maximum of the hourly instantaneous wind speeds over the 
previous twelve hours, given in m\,s$^{-1}$, see e.g. \citet{sgr10}) for
152 stations in the Automated Surface
Observing Network \citep{asos}. The ensemble members are not exchangeable
as they are generated with initial conditions from different sources.

 In the present study we investigate
 forecasts for calendar year 2008 with additional data from the
last month of 2007 used for parameter estimation. After removing
days and locations with missing data 101 stations remain where
the number of days for which forecasts and validating observations are
available varies between 160 and 291.

Figure \ref{fig:fig1}c shows the verification rank histogram of the raw
ensemble where, similar to the previous cases, one can again observe a strongly underdispersive character. The ensemble coverage equals $45.24\,\%$, whereas the nominal coverage for eight ensemble members equals $7/9$, that is $77.78 \,\%$.

\section{Ensemble Model Output Statistics}
   \label{sec:sec3}

As mentioned in the Introduction, the EMOS predictive distribution of a future weather quantity is a single parametric distribution, where the parameters depend on the ensemble. For example, a normal distribution provides a fairly good fit for temperature and pressure \citep{grwg}, whereas wind speed requires a distribution with non-negative support. 

In what follows, we consider three different types of EMOS models: standard EMOS models based on a single parametric family, combination models that select one of multiple parametric distributions based on the values of suitable covariates, and new mixture models. Models based on single parametric families for wind speed which employ truncated normal, log-normal or generalized extreme value distributions, as well as combination models selecting one of these distribution based on covariates have been explored in previous works \citep[see e.g.][]{tg,lt,bl} and are reviewed in Section \ref{subs:subs3.1}. As an alternative choice, we propose new mixture models based on a weighted mixture of truncated normal and log-normal distributions in Section \ref{subs:subs3.2}. The basic EMOS models from previous studies are used as benchmark models in order to assess the predictive performance of the novel mixture models in Section \ref{sec:sec4}.

\subsection{Basic EMOS models}
   \label{subs:subs3.1}

\subsubsection*{Models based on single parametric distributions}
   
The EMOS model introduced by \citet{tg} is based on a truncated normal (TN) distribution, i.e., the predictive distribution is
\begin{equation}
   \label{eq:eq3.1}
  {\mathcal N}_0\big(a_0+a_1f_1+ \cdots +a_Mf_M,b_0+b_1S^2\big) \qquad
  \text{with} \qquad S^2:=\frac 1{M-1}\sum_{k=1}^M\big (f_k-\overline f\big)^2,
\end{equation}
where \ $f_1,f_2,\ldots ,f_M$ \ denote the ensemble of
distinguishable forecasts of wind speed for a given location
and time, \ $\overline f$ \ stands for the ensemble mean, and 
\ ${\mathcal N}_0\big(\mu,\sigma^2\big)$ \ denotes the TN
distribution with location \ $\mu$, \ scale \ $\sigma>0$, \ and 
cut-off at zero having probability density function (PDF) 
\begin{equation*}
g(x\vert\, \mu,\sigma):=\frac{\frac
  1{\sigma}\varphi\big((x-\mu)/\sigma\big)}{\Phi\big(\mu/\sigma\big)
}, \quad x\geq 0, \qquad \text{and} \qquad g(x\vert\, \mu,\sigma):=0,
\quad \text{otherwise,}
\end{equation*}
where \ $\varphi$ \ and \ $\Phi$ \ are the PDF and the cumulative distribution function (CDF) of the
standard normal distribution, respectively. 

As an alternative to the TN distribution \citet{bl} propose the use of a log-normal (LN) distribution where the mean \ $m$ \ and  variance \ $v$ \ are affine functions of the ensemble members and ensemble variance, respectively, i.e., 
\begin{equation}
\label{eq:eq3.2}
m=\alpha_0+\alpha_1f_1+ \cdots +\alpha_Mf_M \qquad\text{and}\qquad
v=\beta_0+\beta_1 S^2.
\end{equation}
Usually the PDF of the LN distribution \ $\mathcal{LN}\big(\mu,\sigma\big)$ 
\ is expressed using location \ $\mu$ \ and shape \ $\sigma>0$ \ parameters and has the form
\begin{equation*}
h(x\vert\, \mu,\sigma):=\frac
  1{x\sigma}\varphi\big((\log x-\mu)/\sigma\big), \quad x\geq 0,
  \qquad \text{and} \qquad h(x\vert\, \mu,\sigma):=0, 
\quad \text{otherwise,}
\end{equation*}
however, it can be easily expressed in terms of mean and variance with the help of transformations
\begin{equation}
  \label{eq:eq3.3}
\mu =\log \bigg(\frac {m^2}{\sqrt{v+m^2}}\bigg) \qquad \text{and}
\qquad \sigma=\sqrt{\log \Big (1+ \frac v{m^2}\Big)}.
\end{equation}

\citet{lt} propose an EMOS approach based on a generalized extreme value (GEV) distribution, however, this model has the disadvantage of assigning positive probability to negative wind speed values \citep[see, e.g.,][]{bl}. 

Location  and scale/shape parameters  of models \eqref{eq:eq3.1} and \eqref{eq:eq3.2} can be estimated from the training data consisting of ensemble members and verifying observations from the preceding \ $n$ \ days, by optimizing an
appropriate verification score (see Section \ref{subs:subs3.3}). 

EMOS models \eqref{eq:eq3.1}  and \eqref{eq:eq3.2} are valid only in the cases when the sources of the ensemble members are clearly distinguishable, which is the case for the UWME described in Section \ref{subs:subs2.3} or for the
Consortium for Small-scale Modeling Germany (COSMO-DE) ensemble of the German Meteorological Service \citep{gtpb}. 
However, in most of the currently used EPSs some members are obtained with the help of perturbations of the initial conditions. These members are statistically indistinguishable and can be considered as exchangeable. This the case for the ECMWF and ALADIN-HUNEPS ensembles described in 
Sections \ref{subs:subs2.1} and \ref{subs:subs2.2}, respectively.

Suppose we have \ $M$ \ ensemble members divided into \ $m$ \ exchangeable
groups, where the \ $k$th \ group contains \ $M_k\geq 1$ \ ensemble
members such that \ $\sum_{k=1}^mM_k=M$. \  In such
situations the ensemble members within an exchangeable group should share the same parameters \citep{gneiting14} resulting in a TN model
\begin{equation}
   \label{eq:eq3.4}
  {\mathcal N}_0\bigg(a_0+a_1\sum_{\ell_1=1}^{M_1}f_{1,\ell_1}+ \cdots
  +a_m\sum_{\ell_m=1}^{M_m} f_{m,\ell_m},b_0+b_1S^2\bigg), 
\end{equation}
and a LN model with mean and variance
\begin{equation}
  \label{eq:eq3.5}
m=\alpha_0+\alpha_1\sum_{\ell_1=1}^{M_1}f_{1,\ell_1}+ \cdots
  +\alpha_m\sum_{\ell_m=1}^{M_m} f_{m,\ell_m} \qquad\text{and}\qquad
  v=\beta_0+\beta_1 S^2,
\end{equation}
where \ $f_{k,\ell}$ \ denotes the  $\ell$th member of the $k$th group.

\subsubsection*{Combination models}

LN and GEV distributions have heavier upper tails than the TN distribution and are therefore more appropriate to model high wind speed values. To combine this advantage with the good performance of the TN model for low and medium wind speeds \citet{lt} and \citet{bl} also examined a regime-switching combination method, where either a TN or a heavy-tail distribution is used depending on the median value of the ensemble forecast. If the ensemble median is below a given threshold, wind speed is modeled by a TN distribution, otherwise a GEV or LN distribution is employed. The optimal threshold value for a given EPS is determined during a preliminary study and it is then fixed over the whole data set. The problem with this approach is that the threshold parameter is static (rarely updated) and cannot adapt to the changes in the ensemble. \citet{bl} also consider a more adaptive method where the threshold parameter is re-estimated as a fixed quantile of the ensemble medians in the corresponding training period for 
each forecast date. However, this approach is computationally more demanding without yielding a significant improvement in predictive performance.

EMOS models based on combining two parametric families by exclusively selecting one of them at each forecast instance also suffer from the drawback that a suitable covariate has to be chosen as a selection criterion. This necessary step limits the flexibility of the combination models in practice as the adequacy of covariates might depend on the data set at hand. While the ensemble median works reasonably well in the data sets considered in this article, this observation might change for different EPSs.

\subsection{Mixture models}
   \label{subs:subs3.2}

In order to combine the advantages of lighter and heavier-tailed distributions and to avoid the aforementioned problems in the process, we introduce new EMOS models based on weighted mixtures of two parametric distributions.

In particular, we propose to model wind speed with a weighted mixture of models \eqref{eq:eq3.1} and \eqref{eq:eq3.2} (or \eqref{eq:eq3.4} and \eqref{eq:eq3.5} for exchangeable ensemble members) resulting in the predictive PDF
\begin{equation}
   \label{eq:eq3.6}
\psi(x\vert\, \mu_{TN},\sigma_{TN};\mu_{LN},\sigma_{LN};\omega):=\omega g(x\vert\, \mu_{TN},\sigma_{TN})+(1-\omega)h(x\vert\, \mu_{LN},\sigma_{LN}),
\end{equation}
where the dependence of parameters \ $\mu_{TN},\ \sigma_{TN}$ \ and \ $\mu_{LN},\ \sigma_{LN}$ \ on the ensemble are given by 
\eqref{eq:eq3.1} (or \eqref{eq:eq3.4}) and \eqref{eq:eq3.2} (or \eqref{eq:eq3.5}) and \eqref{eq:eq3.3}, respectively. In case of model \eqref{eq:eq3.6} location and scale/shape parameters of the TN and LN models together with the weight \ $\omega \ \in [0,1]$ \ are estimated simultaneously by optimizing some verification score over the training data.

Note that instead of a LN distribution, in \eqref{eq:eq3.6} one can incorporate  other non-negative laws with heavy right tails. A natural choice would be the generalized Pareto distribution (GPD) used in extreme value theory \citep[see, e.g.,][]{bf}, however, tests for the ensemble forecasts considered here indicate a worse predictive performance of the TN-GPD model compared with the TN-LN mixture and the benchmark models. 

In comparison with the basic EMOS models proposed in previous work, the new mixture models exhibit desirable properties from a theoretical perspective as they do not require the exclusive choice of one of multiple parametric families and are more flexible than models based on single parametric distributions. Their advantages from a practical perspective such as a significantly improved calibration will be demonstrated in Section \ref{sec:sec4}.

\subsection{Verification scores}
   \label{subs:subs3.3}
The main aim of probabilistic forecasting is to access the maximal sharpness of the predictive distribution subject to calibration \citep{gbr}. Calibration means a statistical consistency between the predictive distributions and the validating observations whereas sharpness refers to the concentration of the predictive distribution. A straightforward way to check the calibration of a probabilistic forecast is the use of probability integral transform (PIT) histograms. The PIT is defined as the value of the predictive CDF evaluated at the verifying observations \citep{rgbp} and the closer the histogram to the uniform distribution, the better the calibration. PIT histograms are continuous analogues of verification rank histograms, a comparison of these histograms can thus be used as a measure of the possible improvements due to statistical post-processing.

Apart from the visual inspection of PIT histograms, formal statistical test of uniformity can be used to assess calibration. As the PIT values of multi-step ahead probabilistic forecast exhibit serial correlation \citep[see, e.g.,][]{dgt98} and the probabilistic forecasts cannot be assumed to be independent in space and time, we employ a moment-based test of uniformity proposed by \citet{knueppel15} which accounts for dependence in the PIT values. In particular, we use the $\alpha_{1234}^0$ test of \citet{knueppel15} that has been demonstrated to have superior size and power properties compared to alternative choices. Due to the large sample size in case of the ECMWF and UWME data, the null hypothesis of uniformity is rejected for all post-processing models. However, as our focus lies on the comparative assessment of calibration, we report bootstrap estimates of the rejection rates of the $\alpha_{1234}^0$ test based on 10\,000 random samples of size 2\,500 each. If a model exhibits superior 
calibration, the null hypothesis of uniformity should be rejected in fewer cases compared to a model with inferior calibration.

Another approach to assess calibration is the investigation of the coverage of the \ $(1-\alpha)100 \,\%$,  $\alpha \in (0,1),$ \ central prediction interval, defined as the proportion of validating observations located between
the lower and upper \ $\alpha/2$ \ quantiles of the predictive
distribution, where \ $\alpha$ \ is chosen to match the nominal coverage of the raw ensemble (ECMWF: $96.08\,\%$; ALADIN-HUNEPS: $83.33\,\%$; UWME: $77.78\,\%$). The coverage of a calibrated predictive PDF should be
around \ $(1-\alpha)100 \,\%$ \ and the proposed choices of \ $\alpha$
\ allow direct comparisons with the raw ensembles. Further, the average widths of these central prediction intervals provide information about the sharpness of the predictive distributions.

Calibration and sharpness can also be addressed simultaneously with the help of scoring rules which measure the predictive performance by numerical values assigned to pairs of probabilistic forecasts and observations \citep{grjasa}. The most popular scoring rules are the logarithmic score (LogS), that is the negative logarithm of the predictive PDF \ $f(y)$ \ evaluated at the verifying observation,
\[
 \logs(F,x) = - \log(f(x)),
\]
and the continuous ranked probability score \citep[CRPS;][]{grjasa,wilks}. The CRPS of a CDF \ $F(y)$ \ and an observation \ $x$ \ is defined as
\begin{equation}
  \label{eq:eq3.7}
\crps\big(F,x\big):=\int_{-\infty}^{\infty}\big (F(y)-{\mathbbm 
  1}_{\{y \geq x\}}\big )^2\,{\mathrm d}y=\int_{-\infty}^x F^2(y){\mathrm d} y +\int_x^{\infty}\big (1-F(y)\big )^2{\mathrm d}y, 
\end{equation}
where \ ${\mathbbm 1}_H$ \ denotes the indicator of a set \ $H$. \
While both the CRPS and the logarithmic score are proper scoring rules \citep{grjasa}, the CRPS can be expressed in the same unit as the observation.

Further, point forecasts such as EMOS and ensemble medians and means are evaluated with the help of mean absolute errors (MAEs) and root mean squared errors (RMESs). We remark that the former is optimal for the median, whereas the latter is for the mean \citep{gneiting11}.

Finally, to evaluate the goodness of fit of probabilistic forecasts to high wind speed values a useful tool to be considered is the threshold-weighted continuous ranked probability score (twCRPS) 
\begin{equation*}
\twcrps\big(F,x\big):=\int_{-\infty}^{\infty}\big (F(y)-{\mathbbm 
  1}_{\{y \geq x\}}\big )^2\omega(y){\mathrm d}y
\end{equation*}
introduced by \citet{gr}, where \ $\omega(y)\geq 0$ \ is a weight
function. Obviously, \ $\omega(y)\equiv 1$ \ yields the
traditional CRPS defined by  \eqref{eq:eq3.7}, while one may set \ $\omega
(y)={\mathbbm 1}_{\{y \geq r\}}$\  to address wind
speeds above a given threshold \ $r$. \  Similar to \citet{lt} and \citet{bl}, where the upper tail behaviors of regime-switching EMOS models are investigated,
we consider threshold values approximately corresponding to the 90th, 95th and
99th percentiles of the wind speed observations. One can also quantify the improvement in twCRPS with respect to some reference predictive
CDF \ $F_{ref}$ \ with the help of the threshold-weighted continuous
ranked probability skill score  \citep[twCRPSS; see,
e.g.,][]{lt} defined as
\begin{equation*}
\twcrpss\big(F,x\big):=1-\frac{\twcrps\big(F,x\big)}{\twcrps\big(F_{ref},x\big)}.
\end{equation*}
This score is obviously positively oriented, and in this study
the predictive CDF corresponding to the classical TN model is used as a reference.

In order to assess the statistical significance of observed score differences between the models we use formal statistical tests of equal predictive performance. Diebold-Mariano \citep[DM;][]{dm95} tests allow to account for dependence in the forecast errors and are widely used in the econometric literature. Denote the mean values of a proper scoring rule \ $S$ \  for two competing probabilistic forecasts \ $F_t$ \  and \ $G_t$ \  by \ $\bar S(F_t,x_t) = \frac{1}{N}\sum_{t\in T} S(F_t, x_t)$ \  and \ $\bar S(G_t, x_t) \frac{1}{N}\sum_{t\in T} S(G_t, x_t)$, \  respectively, where \ $t\in T$ \  denotes the forecast cases in a test set \ $T$ \ of size \ $N$. \ The test statistic of the DM test is given by
\begin{equation}\label{eq:DM}
  t_N = \sqrt{N} \frac{\bar S(F_t,x_t) - \bar S(G_t, x_t)}{\hat \sigma_N}, 
\end{equation}

where \ $\hat \sigma_N$ \ is a suitable estimator of the asymptotic standard deviation of the sequence of score differences \ $S(F_t,x_t) - S(G_t, x_t)$. \ Under some weak regularity assumptions, \ $t_N\ $ asymptotically follows a standard normal distribution under the null hypothesis of equal predictive performance. Negative values of \ $t_N$ \ indicate a better predictive performance of \ $F$, \  whereas \ $G$ \  is preferred in case of positive values of  \ $t_N$. \  The statistical significance of the observed values of the test statistic can be assessed by computing the corresponding p-values under the null hypothesis. 
Following suggestions of \citet{dm95} and \citet{gr}, we estimate the autocovariance of the sequence of score differences in \eqref{eq:DM} by the sample autocovariance  up to lag $h-1$ in case of $h$ step ahead forecasts. The results of DM tests based on the LogS, the CRPS and the twCRPS are discussed in Section \ref{sec:sec4} for the individual ensembles.

\section{Results}
  \label{sec:sec4}

As mentioned in Section \ref{sec:sec1}, the predictive skills of the mixture model \eqref{eq:eq3.6} are tested on the 50 member ECMWF ensemble, the ALADIN-HUNEPS ensemble of the HMS and the eight-member UWME. The three EPSs differ both in generation of the ensemble members and in the predicted wind speed quantity. 
Model performances are evaluated with the help of the verification scores given in Section \ref{subs:subs3.3}.
The basic TN, LN and TN-LN regime-switching combination EMOS models proposed in previous studies are used as benchmark models to assess the predictive performance of the new mixture models. For a detailed evaluation  and comparison of the different basic EMOS models, we refer to \citet{bl}. We further  compare the forecasts based on post-processing with the raw ensemble and with climatological forecasts where the observations of the training period are considered as an ensemble.

Following the ideas of \citet{grwg} and \citet{tg} the parameters of the TN, LN and TN-LN mixture models are estimated by minimizing the mean CRPS of the predictive CDFs and corresponding validating observations over the training period. 
However, in case of model \eqref{eq:eq3.6} the CRPS can be evaluated only numerically, resulting in very long optimization procedures. Numerical minimization of the right-most part of equation \eqref{eq:eq3.7}, i.e., 
\ $\crps\big(F,x\big) = \int_{-\infty}^x F^2(y){\mathrm d} y +\int_x^{\infty}\big (1-F(y)\big )^2{\mathrm d}y$ \  leads to slightly lower computation times and better verification scores compared to minimizing alternative representations of the CRPS integral.

Due to the large computational costs of minimum CRPS estimation, we also investigate maximum likelihood (ML) estimation of the parameters. ML estimation corresponds to minimizing the mean logarithmic score which has a simple and closed form. From a theoretical statistics perspective, both estimation approaches fit into a general framework of optimum score estimation and share asymptotic properties such as consistency, see \citet{grjasa} for details. In applications to post-processing ensemble forecasts, the CRPS is often seen as the more appropriate scoring rule for parameter estimation due to the lower sensitivity 
to outliers and extreme events compared to the LogS \cite[see, e.g.,][]{grwg,tg}. In figures and tables, the corresponding mixture models are denoted by TN-LN mix. (CRPS) and TN-LN mix. (ML).

Finally, in order to ensure the comparability with the benchmark models the same training period lengths and for the TN-LN regime-switching models the same thresholds and parameter estimation techniques as in \citet{bl} are employed.

\subsection{ECMWF ensemble}
   \label{subs:subs4.1}

\begin{figure}[t]
\begin{center}
\leavevmode
\epsfig{file=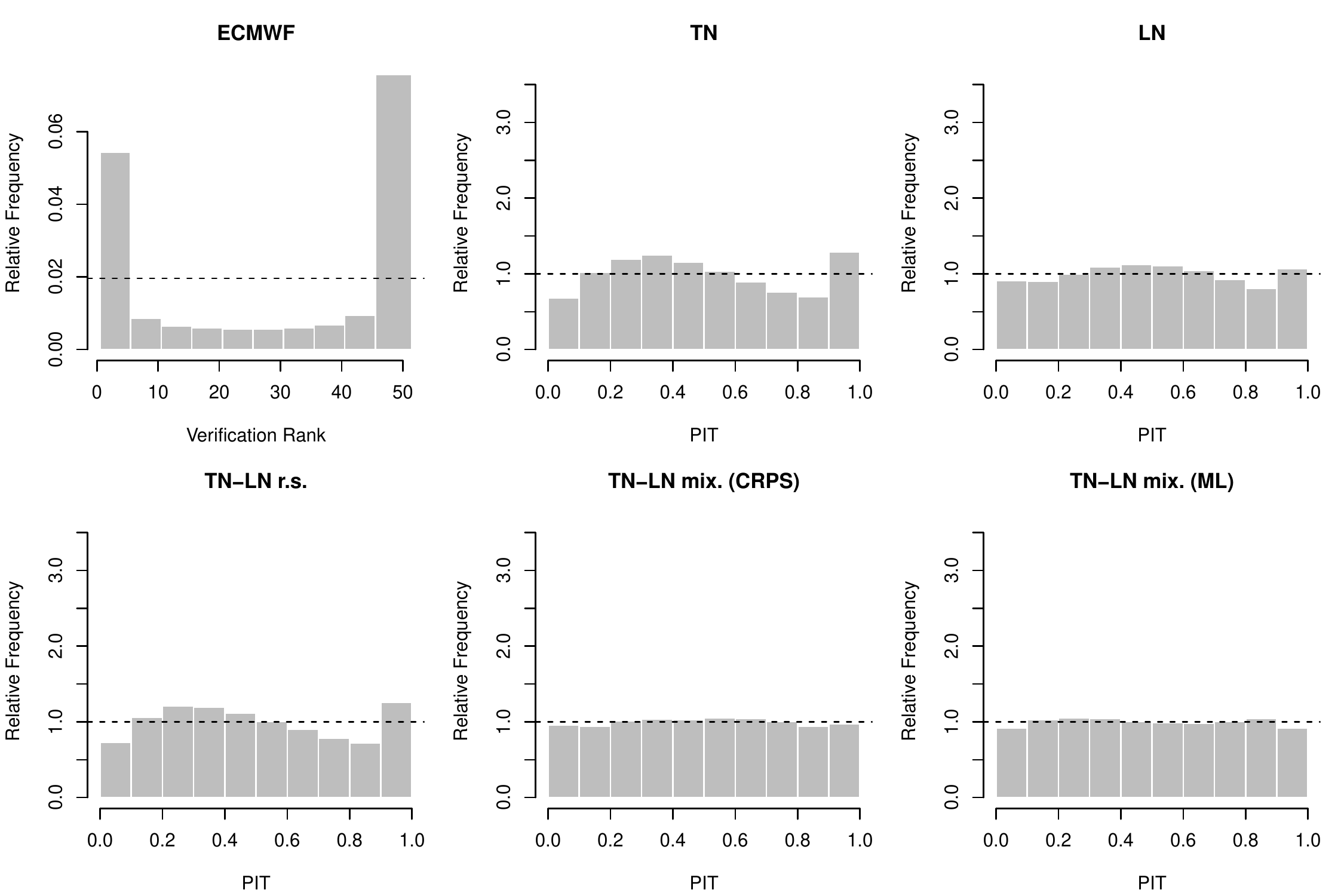,width=16.5cm, angle=0}
\end{center}
\caption{Verification rank histogram of the raw ensemble and PIT
  histograms of the EMOS post-processed forecasts for the ECMWF ensemble.} 
\label{fig:fig2}
\end{figure}

As the fifty members of the ECMWF ensemble are fully exchangeable, the dependencies of the parameters of the TN and LN distributions on the ensemble members are specified by \eqref{eq:eq3.4} and \eqref{eq:eq3.5}, respectively, with \ $m=1$ \ and \ $M=M_1=50$. \ 

The preliminary study by \citet{bl} suggested that the optimal training period length for this particular data set is 20 days, whereas the optimal value of the threshold parameter \ $\theta$ \ of the TN-LN regime-switching combination model equals 8 m\,s$^{-1}$, resulting in the use of an LN distribution in about $14\,\%$ of the forecast cases. As mentioned in Section \ref{subs:subs2.1}, model verification is performed on 83\,220 forecast cases from the one year period between 1 May 2010 and 30 April 2011.

Figure \ref{fig:fig2} showing the verification rank histogram of the raw ensemble and the PIT histograms of the investigated EMOS models clearly indicates that statistical post-processing significantly improves the calibration of the raw ensemble. However, the histograms of TN and TN-LN regime-switching models are still biased, to a smaller extent the same applies for the LN model, whereas the PIT values of mixture model \eqref{eq:eq3.6} with both parameter estimation methods suggest a better fit to the desired uniform distribution.

\begin{table}[t!]
 \begin{center}
  \caption{Bootstrap estimates of rejection rates of the $\alpha_{1234}^0$ test of uniformity based on 10\,000 random samples of size 2\,500 each at the 0.05 level for the different data sets. Lower rejection rates correspond to better calibrated forecasts with the null hypothesis of uniformity being rejected on fewer occasions.} \label{tab:RR_PIT_tests}
  \vskip .5 truecm
  \begin{tabular}{lccccc}
   Ensemble & TN & LN & TN-LN r.s. & TN-LN mix. (CRPS) & TN-LN mix. (ML) \\ \hline 
   ECMWF    & 1  & 1  &  1  & 0.68 & 0.25 \\
   ALHU     & 1  & 1  &  1  & 0 & 0.01 \\
   UWME     & 1  & 1  & 0.68 & 0.47 & 0.31 \\
  \end{tabular}
 \end{center}
\end{table}

In order to quantify the observed differences in calibration, Table \ref{tab:RR_PIT_tests} shows rejection rates of the $\alpha_{1234}^0$ test of uniformity based on random sub-samples. It can be observed that for the ECMWF data, the null hypothesis of uniformity is rejected in all of the cases for the TN, LN and TN-LN combination model, whereas the novel mixture models show much lower rejection rates. This observation is clearly in line with the visual inspection of the PIT histograms in Figure \ref{fig:fig2}.

The positive effect of post-processing can also be observed in Table \ref{tab:tab1} summarizing the verification scores for different probabilistic forecasts together with the average width and coverage of the $96.08\,\%$  central prediction intervals. 
The improvement with respect to the raw ensemble and climatology is quantified in lower CRPS, twCRPS, MAE and RMSE values and the EMOS predictive PDFs result in calibrated central prediction intervals with coverages very close to the nominal value. The much wider central prediction intervals of the EMOS models compared to the ensemble are a natural consequence of the underdispersive character of the latter.

\begin{table}[t!]
\begin{center}
\caption{Mean CRPS, mean twCRPS for various thresholds \ $r$, \ MAE of
  median and RMSE of mean forecasts and coverage and 
  average width of $96.08\,\%$ central prediction intervals
  for the ECMWF ensemble.} \label{tab:tab1} 

\vskip .5 truecm
\begin{tabular}{lcccccccc} 
Forecast&CRPS&\multicolumn{3}{c}{
  twCRPS m\,s$^{-1}$}&MAE&RMSE&Cover.&Av.w.\\\cline{3-5}
&m\,s$^{-1}$&$r\!=\!10$&$r\!=\!12$&$r\!=\!15$&
m\,s$^{-1}$&m\,s$^{-1}$&$(\%)$&m\,s$^{-1}$\\ \hline
TN-LN mix. (CRPS)&1.030&0.194&0.106&0.041&1.384&2.135&94.34&7.71\\
TN-LN mix. (ML)&1.034&0.196&0.108&0.041&1.391&2.138&95.81&8.72\\
TN&1.045&0.200&0.110&0.042&1.388&2.148&92.19&6.39 \\
LN&1.037&0.198&0.109&0.042&1.386&2.138&93.16&6.91 \\
TN-LN r.s. ($\theta\!=\!8.0$)&1.033&0.191&0.103&0.039&1.379&2.135&
92.49&6.36 \\ \hline 
Ensemble&1.263&0.211&0.113&0.043&1.441&2.232&45.00&1.80 \\ 
Climatology&1.550&0.251&0.128&0.045&2.144&2.986&
95.84&11.91 
\end{tabular} 
\end{center}
\end{table}

\begin{figure}[t!]
\begin{center}
\leavevmode
\epsfig{file=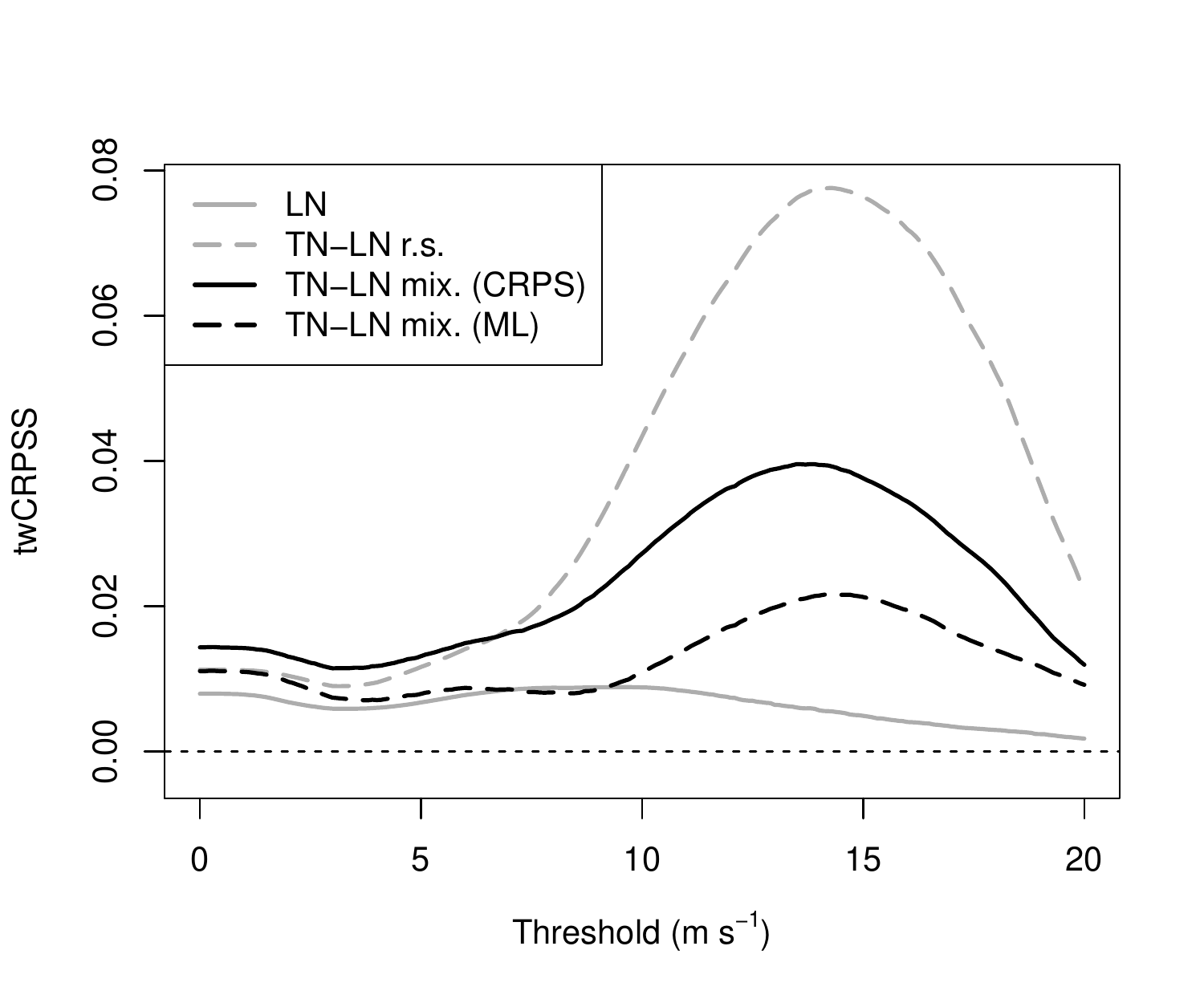,height=7.3cm, angle=0}
\caption{twCRPSS values for the ECMWF ensemble with TN as reference model.}   
\label{fig:fig3}
\end{center}
\end{figure}

Among the competing post-processing methods the TN-LN mixture and regime-switching models clearly outperform the TN and LN EMOS approaches in almost all scores investigated. The lowest CRPS value belongs to the mixture model with parameters estimated by optimizing the mean CRPS, whereas the regime-switching approach produces the best MAE, RMSE and twCRPS scores. The two parameter estimation methods make only a very slight difference in model performance (ML estimation leads to slightly worse scores) and the TN-LN  mixture EMOS models are fully able to keep up with the regime-switching approach. This ranking of models can also be observed in Figure \ref{fig:fig3} displaying the twCRPSS values of the LN, TN-LN regime-switching and TN-LN mixture (with CRPS and logarithmic score optimization) EMOS methods with respect to the reference TN EMOS model as functions of the threshold \ $r$. \ While the regime-switching approach outperforms the other models for all threshold values and has the best overall performance, it is clearly less flexible than the mixture models which show the second best performance.

\begin{table}
 \begin{center}
  \caption{Values of the test statistics $t_N$ of the two-sided DM test of equal predictive performance \eqref{eq:DM} for the comparison of the TN-LN mixture models and the benchmark models for the ECMWF ensemble. Negative values indicate a superior predictive performance of the mixture model in the left column, and positive values indicate a better performance of the benchmark model. 
  Values that are significant at the 0.05 level under the null hypothesis of equal predictive performance are printed in bold.} \label{tab:DM-ECMWF}
  
  \vskip .5 truecm
  \begin{tabular}{lccc}
    & TN & LN & TN-LN r.s. \\ \hline 
   \multicolumn{4}{l}{\textit{DM test based on LogS}} \\ 
   \hline
   TN-LN mix. (CRPS) & \textbf{-36.67} & \textbf{-24.25} & \textbf{-32.79} \\
   TN-LN mix. (ML) & \textbf{-36.43} & \textbf{-28.91} & \textbf{-32.86} \\
   \hline
     \multicolumn{4}{l}{\textit{DM test based on CRPS}} \\ 
   \hline
   TN-LN mix. (CRPS) & \textbf{-45.10} & \textbf{-27.10} & \textbf{-4.71} \\
   TN-LN mix. (ML) & \textbf{-30.29} & \textbf{-11.42} & 0.32 \\
   \hline
     \multicolumn{4}{l}{\textit{DM test based on twCRPS with threshold $r = 10$}} \\ 
   \hline
   TN-LN mix. (CRPS) & \textbf{-21.55} & \textbf{-17.32} & \textbf{6.32} \\
   TN-LN mix. (ML) & \textbf{-10.49} & \textbf{-2.83} & \textbf{9.92} \\
  \end{tabular}
 \end{center} 
\end{table}

Table \ref{tab:DM-ECMWF} summarizes the values of the test statistics of DM tests in order to assess the statistical significance of the observed score differences between the novel mixture models and the benchmark models. It can be observed that in terms of all employed proper scoring rules, the mixture models perform significantly better than the TN and LN models. The results of the comparison with the TN-LN combination model depend on the scoring rule. While the mixture models perform significantly better in terms of the LogS, the combination model is preferred in terms of the twCRPS.
 
\begin{figure}[t!]
\begin{center}
\leavevmode
\epsfig{file=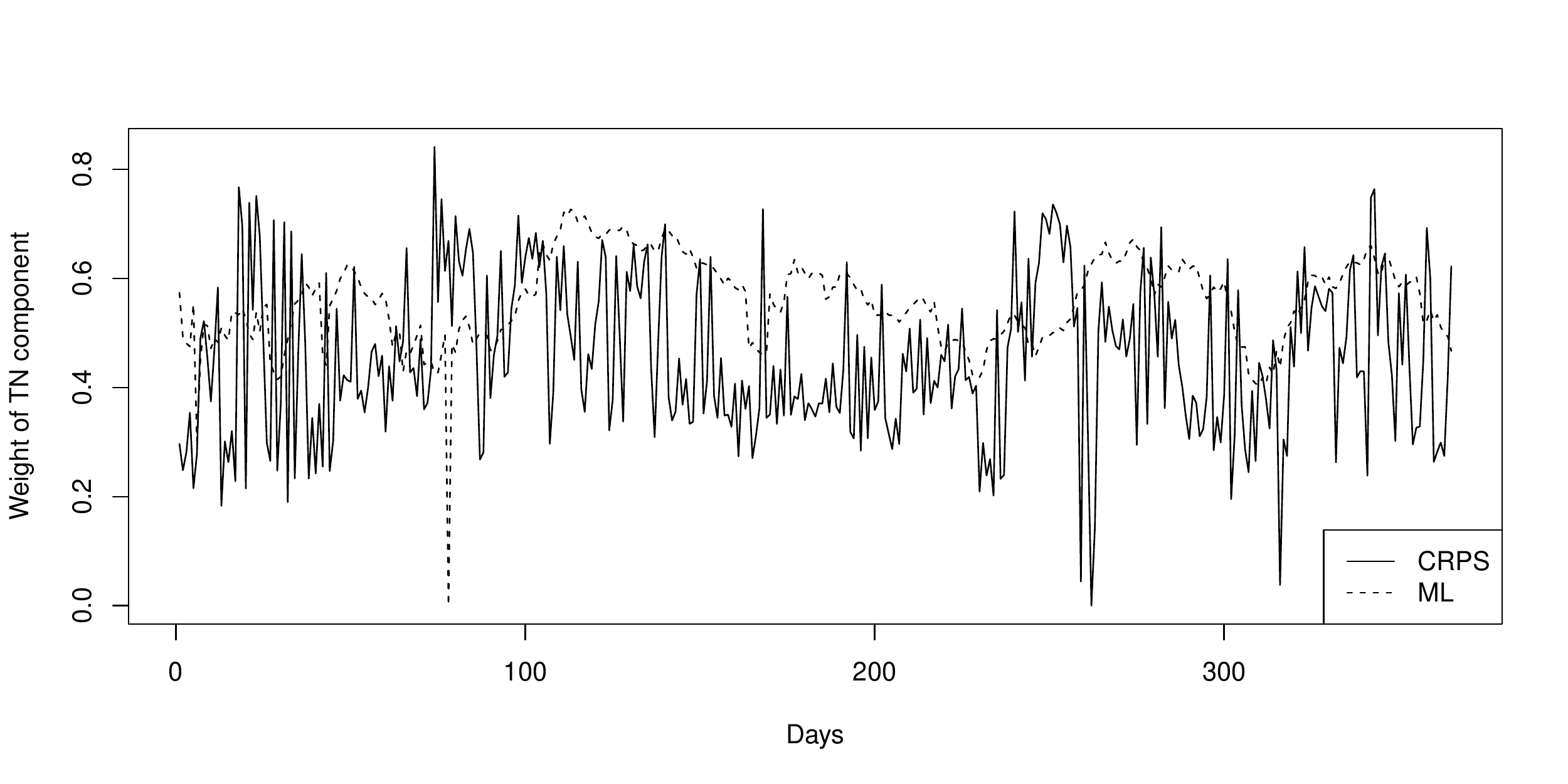,height=7.5cm, angle=0}
\caption{Weights of the TN component  for the ECMWF ensemble.}   
\label{fig:fig4}
\end{center}
\end{figure}

Figure \ref{fig:fig4} shows the weights \ $\omega$ \ of the mixture model \eqref{eq:eq3.6} estimated using optimizations with respect to the mean CRPS and the mean logarithmic score over the training data. Despite the similar predictive skills (see Table \ref{tab:tab1}), the two parameter estimating methods result in completely different sets of weights having only a 
minor non-significant correlation of \ $0.063$. \  However, having a closer look at the predictive PDFs one can observe that the corresponding locations and scales/shapes of the TN and LN components produced by the two different estimation methods are strongly correlated, their correlations vary between $0.921$ and $0.968$, except for the scales of the TN component with a correlation of $0.283$.

\subsection{ALADIN-HUNEPS ensemble}
   \label{subs:subs4.2}

\begin{figure}[t]
\begin{center}
\leavevmode
\epsfig{file=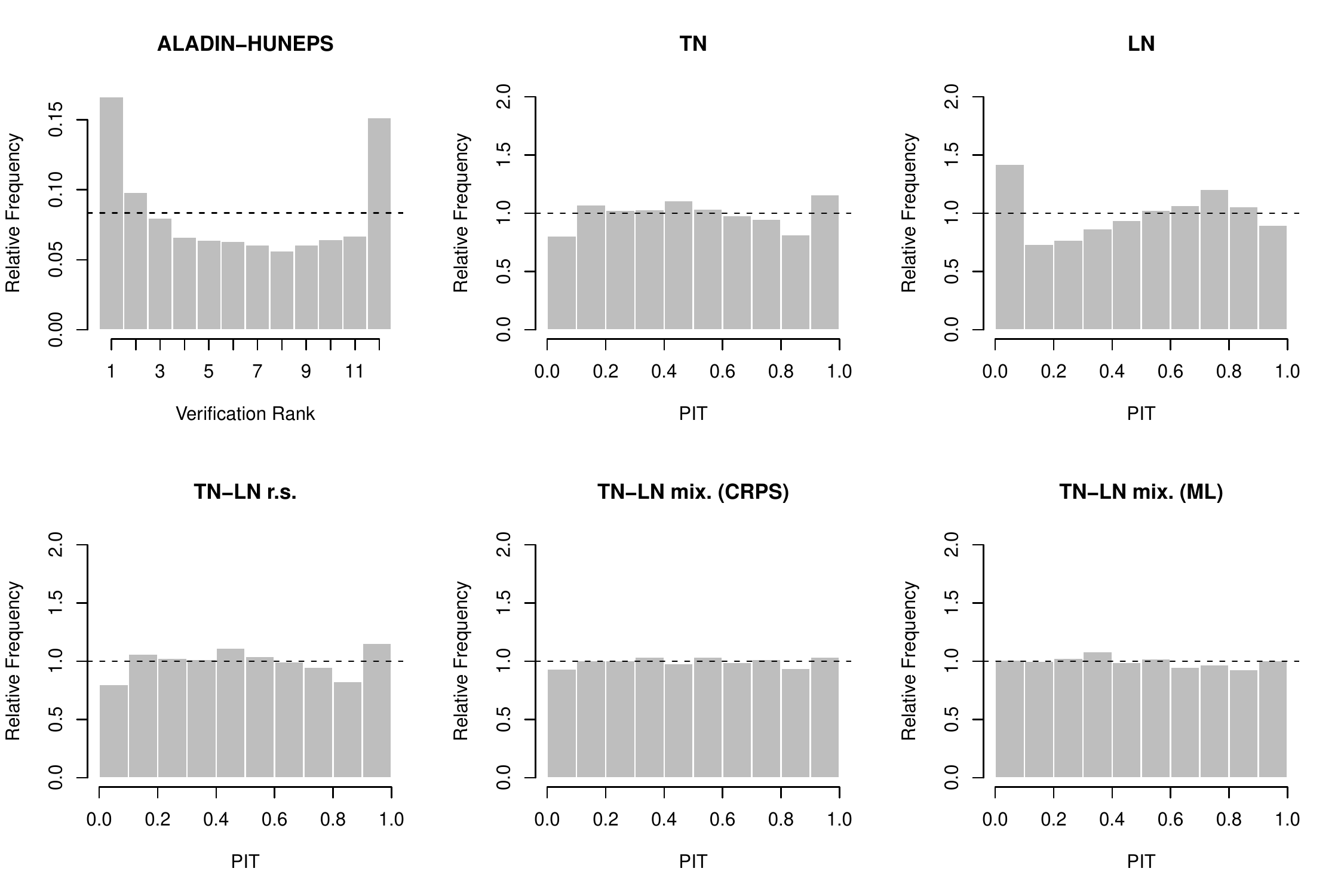,width=16.5cm, angle=0}
\end{center}
\caption{Verification rank histogram of the raw ensemble and PIT
  histograms of the EMOS post-processed forecasts for the ALADIN-HUNEPS ensemble.} 
\label{fig:fig5}
\end{figure}

The ALADIN-HUNEPS ensemble consists of a control member and 10 exchangeable ensemble members (see Section \ref{subs:subs2.2}) inducing a natural splitting of the ensemble into two groups. The first group contains just the control, whereas the second group consists of the 10 exchangeable ensemble members. This results in TN and LN models \eqref{eq:eq3.4} and \eqref{eq:eq3.5}, respectively, where \ $m=2$, \ with \ $M_1=1$ \ and \ $M_2=10$. \ 

For this particular ensemble \citet{bhn2} and \citet{bl} showed that a training period of length 43 days is optimal both for the TN and the LN EMOS models, whereas the optimal threshold for the TN-LN regime-switching combination model is \ $\theta=6.9$ m\,s$^{-1}$. For this threshold value, a LN distribution is used in $4\,\%$ of the forecast cases.

\begin{table}[t!]
\begin{center}
\caption{Mean CRPS, mean twCRPS for various thresholds \ $r$, \ MAE of
  median and RMSE of mean forecasts and coverage and 
  average width of $83.33\,\%$ central prediction intervals 
  for the ALADIN-HUNEPS ensemble.} \label{tab:tab2} 

\vskip .5 truecm
\begin{tabular}{lcccccccc} 
Forecast&CRPS&\multicolumn{3}{c}{
  twCRPS m\,s$^{-1}$}&MAE&RMSE&Cover.&Av.w.\\\cline{3-5}
&m\,s$^{-1}$&$r\!=\!6$&$r\!=\!7$&$r\!=\!9$&
m\,s$^{-1}$&m\,s$^{-1}$&$(\%)$&m\,s$^{-1}$\\ \hline
TN-LN mix. (CRPS)&0.736&0.100&0.053&0.011&1.037&1.358&83.02&3.62\\
TN-LN mix. (ML)&0.737&0.100&0.053&0.012&1.040&1.360&83.14&3.58\\
TN&0.738&0.102&0.054&0.012&1.037&1.357&83.59&3.53 \\
LN&0.741&0.102&0.054&0.011&1.038&1.362&80.44&3.57 \\
TN-LN r.s. ($\theta\!=\!6.9$)&0.737&0.101&0.054&0.011&1.035&
1.356&83.59&3.54\\ 
\hline 
Ensemble&0.803&0.112&0.059&0.013&1.069&1.373&68.22&2.88 \\
Climatology&1.046&0.127&0.064&0.012&1.481&
1.922&82.54&3.43 
\end{tabular} 
\end{center}
\end{table}

Using a 43 days training period one has ensemble forecasts and verifying observations for 315 calendar days (i.e., 3\,150 forecast cases) between 15 May 2012 and 31 March 2013. Figure \ref{fig:fig5} shows the PIT histograms of the various post-processing methods together with the verification rank histogram of the ALADIN-HUNEPS ensemble. Again, compared with the raw ensemble one can clearly see the improvement in calibration of post-processed forecasts, whereas from the competing EMOS methods the two variants of the mixture model \eqref{eq:eq3.6} provide the most uniform PIT histograms.

The significantly better calibration of the mixture models can also be observed from the rejection rates of the $\alpha_{1234}^0$ test reported in Table \ref{tab:RR_PIT_tests}. The null hypothesis of uniformity of the PIT values is not rejected for almost all of the random samples, whereas it is rejected on almost all occasions for the competing models based on single parametric distributions and the TN-LN regime-switching combination model.

\begin{figure}[t!]
\begin{center}
\leavevmode
\epsfig{file=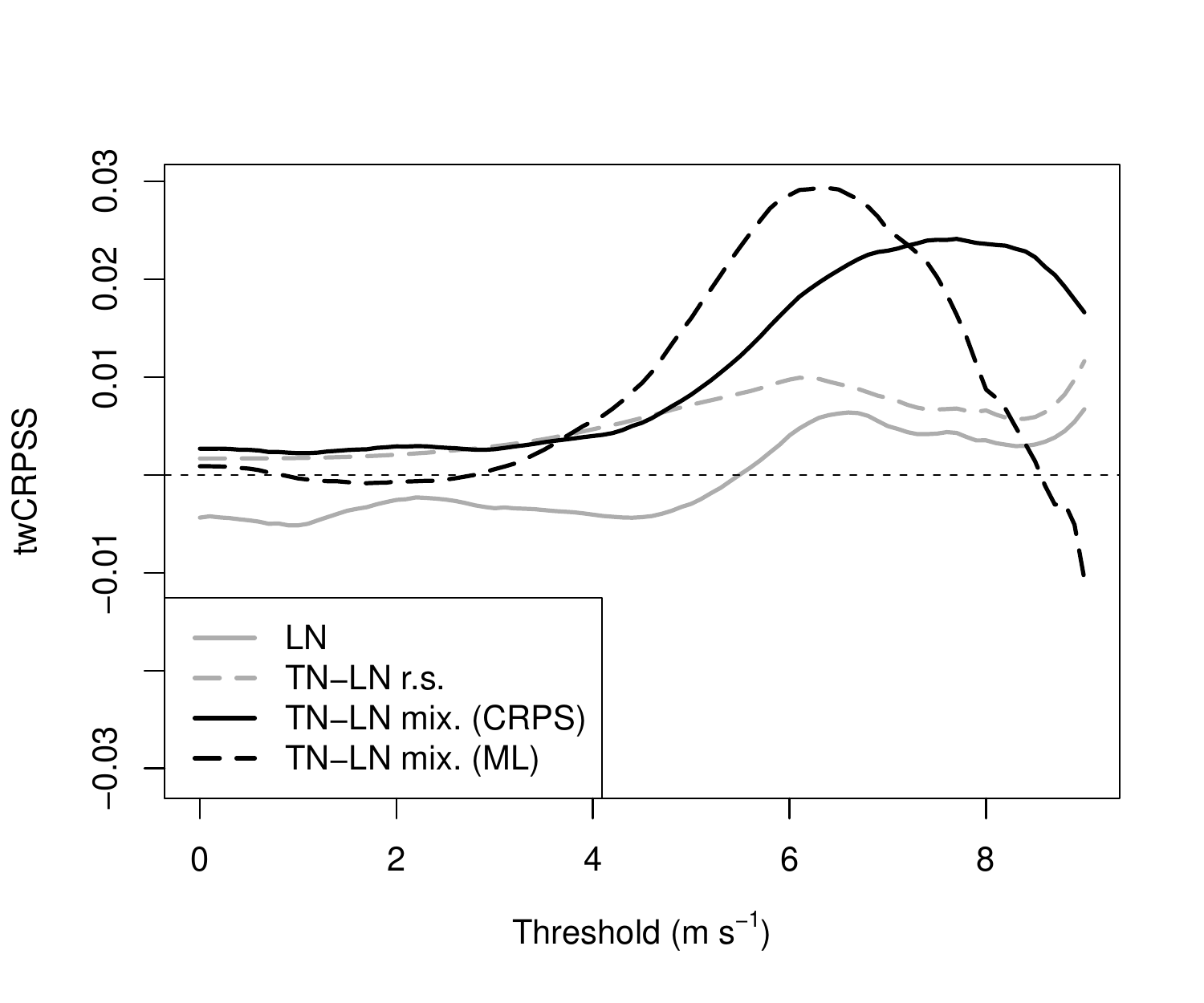,height=7.3cm, angle=0}
\caption{twCRPSS values for the ALADIN-HUNEPS ensemble with TN as reference model.}   
\label{fig:fig6}
\end{center}
\end{figure}

In Table \ref{tab:tab2} the verification scores of probabilistic and point forecasts and coverage and average width of $83.33\,\%$ central prediction intervals are given for the various EMOS models, the ALADIN-HUNEPS ensemble and climatological forecasts. The raw ensemble outperforms climatology and produces sharp forecasts, however, at the cost of being uncalibrated. Post-processing substantially improves the calibration and predictive skill of the raw ensemble, which is in line with the shapes of histograms displayed in Figure \ref{fig:fig5}. The TN-LN mixture and regime-switching combination models show some small improvements over the TN and LN models in terms of all scoring rules and display almost the same predictive performance. For small threshold values the two versions of the mixture model slightly outperform the three benchmark EMOS approaches, however, above the 99th percentile of the validating observations (9 m\,s$^{-1}$), their performances decay quickly, see Figure \ref{fig:fig6}. 

\begin{table}
 \begin{center}
  \caption{Values of the test statistics $t_N$ of the two-sided DM test of equal predictive performance \eqref{eq:DM} for the comparison of the TN-LN mixture models and the benchmark models for the ALHU ensemble. Negative values indicate a superior predictive performance of the mixture model in the left column, and positive values indicate a better performance of the benchmark model. 
  Values that are significant at the 0.05 level under the null hypothesis of equal predictive performance are printed in bold.} \label{tab:DM-ALHU}
  
  \vskip .5 truecm
  \begin{tabular}{lccc}
    & TN & LN & TN-LN r.s. \\ \hline 
   \multicolumn{4}{l}{\textit{DM test based on LogS}} \\ 
   \hline
   TN-LN mix. (CRPS) & 0.71 & \textbf{-4.78} & 0.63 \\
   TN-LN mix. (ML) & \textbf{-2.43} & \textbf{-5.36} & \textbf{-2.09} \\
   \hline
     \multicolumn{4}{l}{\textit{DM test based on CRPS}} \\ 
   \hline
   TN-LN mix. (CRPS) & \textbf{-2.03} & \textbf{-3.73} & -0.71 \\
   TN-LN mix. (ML) & -0.58 & \textbf{-2.56} & 0.31 \\
   \hline
     \multicolumn{4}{l}{\textit{DM test based on twCRPS with threshold $r = 6$}} \\ 
   \hline
   TN-LN mix. (CRPS) & \textbf{-3.19} & -1.90 & -1.14 \\
   TN-LN mix. (ML) & \textbf{-2.24} & -1.86 & -1.11 \\
  \end{tabular}
 \end{center}
\end{table}

In order to assess the statistical significance of the score differences results of the DM tests of equal predictive performance are reported in Table \ref{tab:DM-ALHU}. In terms of LogS and CRPS, the mixture models significantly outperform the LN model, and the same observation holds in terms of the twCRPS for the comparison with the TN model. In comparison with the TN-LN combination model, the preferred model depends on the employed scoring rule. The only significant score difference in this comparison is in terms of the LogS and favors the mixture model based on ML estimation.

\begin{figure}[t!]
\begin{center}
\leavevmode
\epsfig{file=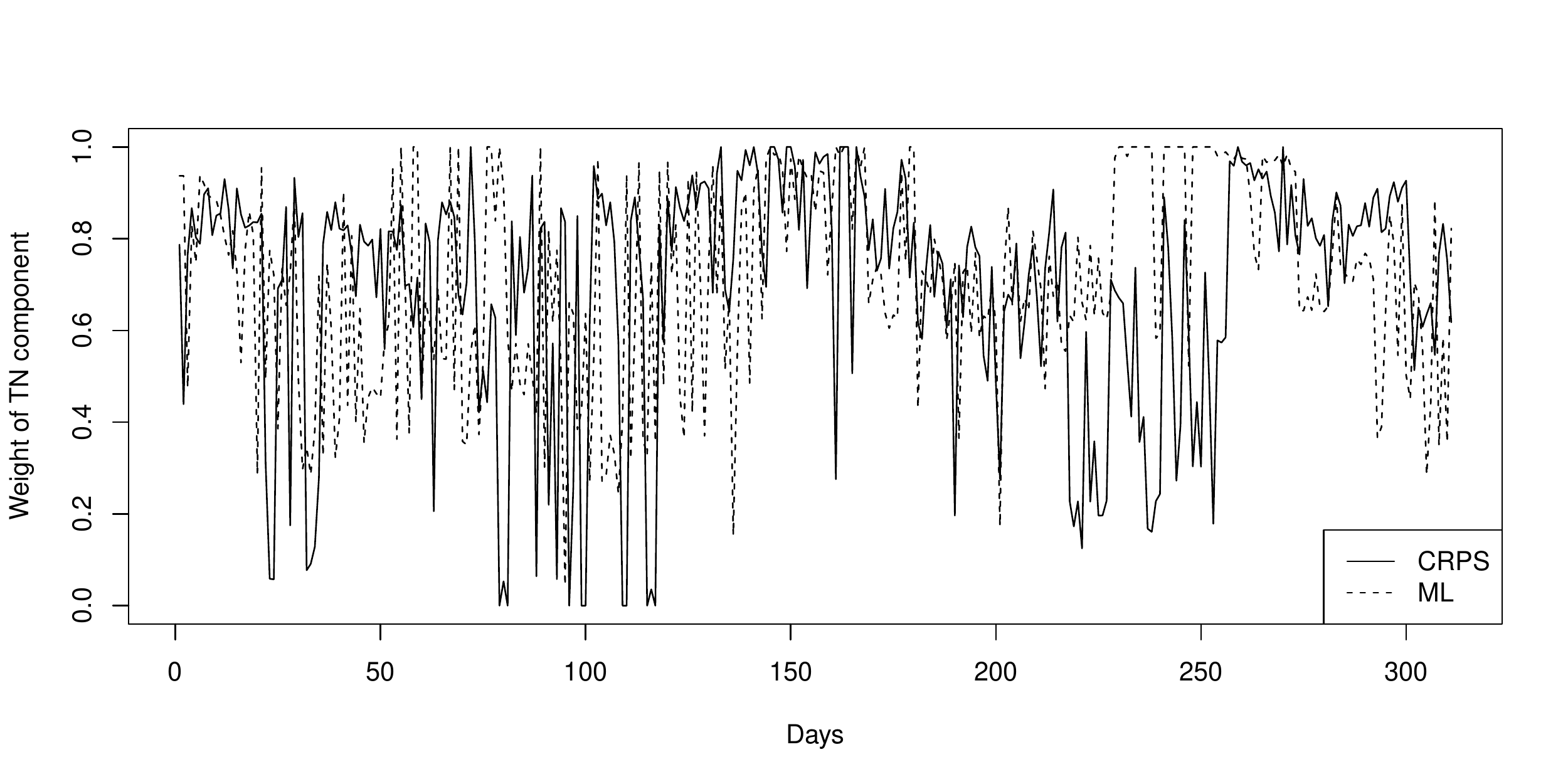,height=7.5cm, angle=0}
\caption{Weights of the TN component for the ALADIN-HUNEPS ensemble.}   
\label{fig:fig7}
\end{center}
\end{figure} 

Finally, similar to the previous case study, the weights belonging to the two parameter estimation methods for the TN-LN mixture model (see Figure \ref{fig:fig7}) are uncorrelated, whereas the correlations of the corresponding location and scale/shape parameters of the TN \ ($\mu_{TN}$ \ and \ $\sigma_{TN}$) \ and LN components \ ($\mu_{LN}$ \ and \ $\sigma_{LN}$) \ are $0.875, \ 0.660$ and $0.747, \ 0.414$, respectively.

\subsection{University of Washington Mesocale Ensemble}
   \label{subs:subs4.3}

The members of the UWME are clearly distinguishable, as they are generated using initial conditions from eight different sources. Hence, location and scale/shape parameters of the TN and LN models are linked to the ensemble via \eqref{eq:eq3.1} and \eqref{eq:eq3.2}, respectively, with \ $M=8$. According to \citet{bl} the optimal training period for this data set is of length 30 days and the optimal threshold value \ $\theta$ \  of the TN-LN combination model is equal to 5.7 m\,s$^{-1}$. Ensemble forecasts for calendar year 2008 are calibrated using these parameters, and in case of the regimes-switching approach a LN model is used in around one third of the 27\,481 individual forecast cases.

\begin{figure}[t]
\begin{center}
\leavevmode
\epsfig{file=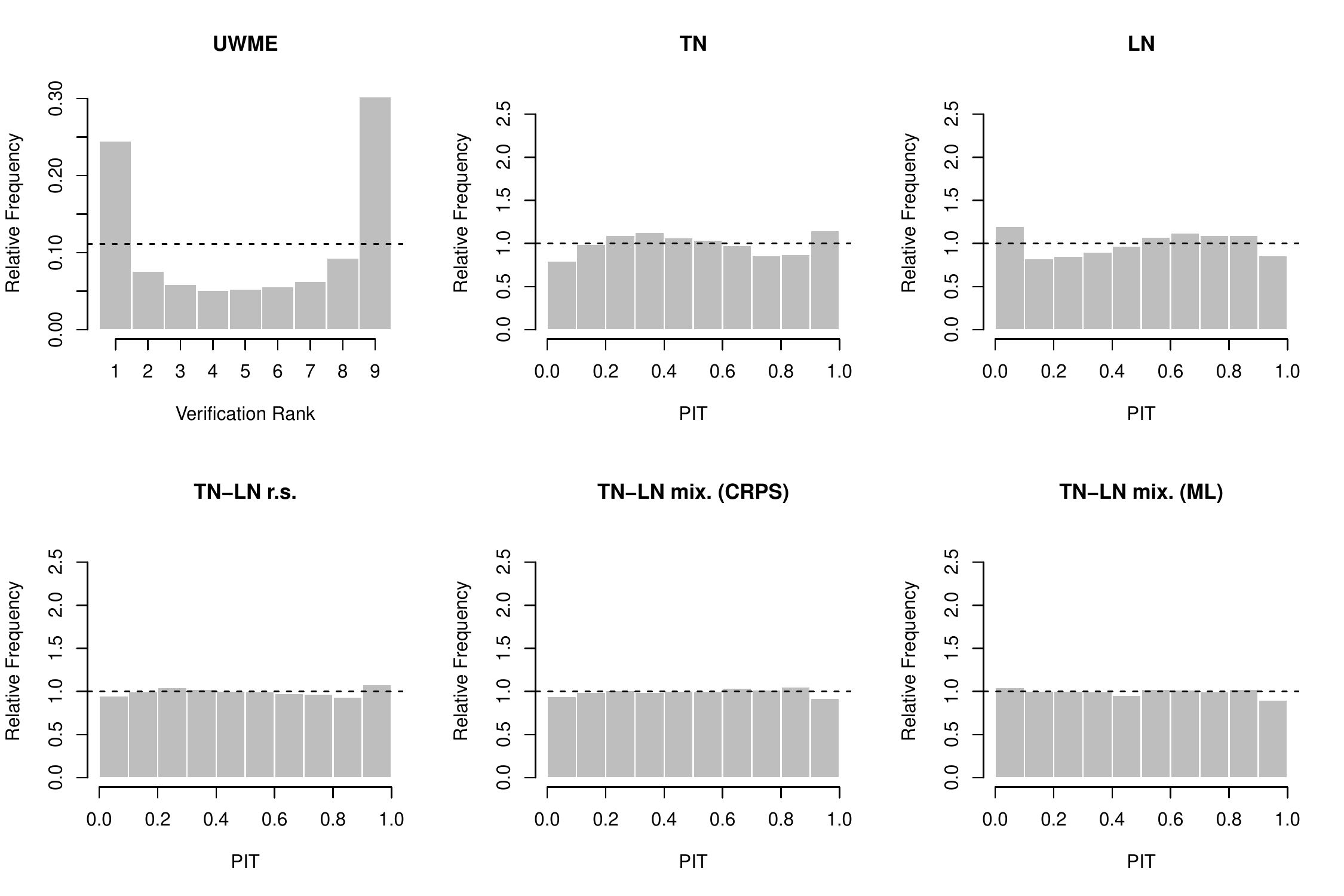,width=16.5cm, angle=0}
\end{center}
\caption{Verification rank histogram of the raw ensemble and PIT
  histograms of the EMOS post-processed forecasts for the UWME.} 
\label{fig:fig8}
\end{figure}

Similar to the previous two sections consider first the PIT histograms of the EMOS predictive distributions displayed in Figure \ref{fig:fig8}. Compared with the verification rank histogram of the
raw ensemble, all post-processing methods result in significant
improvements in the goodness of fit to the uniform distribution,
while, from the various calibration methods, the TN-LN mixture and regime-switch\-ing models have the best performance.

The rejection rates of the $\alpha_{1234}^0$ tests reported in Table \ref{tab:RR_PIT_tests} indicate that the TN-LN mixture model based on ML estimation exhibits the best calibration followed by the mixture model based on minimum CPRS estimation and the TN-LN combination model.

\begin{table}[t!]
\begin{center}
\caption{Mean CRPS, mean twCRPS for various thresholds \ $r$, \ MAE of
  median and RMSE of mean forecasts and coverage and 
  average width of $77.78\,\%$ central prediction intervals
  for the UWME.} \label{tab:tab3}

\vskip .5 truecm
\begin{tabular}{lcccccccc} 
Forecast&CRPS&\multicolumn{3}{c}{
  twCRPS m\,s$^{-1}$}&MAE&RMSE&Cover.&Av. w.\\\cline{3-5}
&m\,s$^{-1}$&$r\!=\!9$&$r\!=\!10.5$&$r\!=\!14$&
m\,s$^{-1}$&m\,s$^{-1}$&$(\%)$&m\,s$^{-1}$\\ \hline
TN-LN mix. (CRPS)&1.105&0.147&0.073&0.010&1.550&2.045&79.02&4.77\\
TN-LN mix. (ML)&1.108&0.147&0.073&0.010&1.560&2.062&78.12&4.78\\
TN&1.114&0.150&0.074&0.010&1.550&2.048&78.65&4.67 \\
LN&1.114&0.147&0.073&0.010&1.554&2.052&77.29&4.69 \\
TN-LN r.s. ($\theta\!=\!5.7$)&1.105&0.149&0.073&0.010&1.550&2.050&77.73&
4.64 \\ \hline 
Ensemble&1.353&0.175&0.085&0.011&1.655&2.169&45.24&2.53 \\
Climatology&1.412&0.173&0.081&0.010&1.987&2.629&81.10&5.90 
\end{tabular}  
\end{center}
\end{table}
  
Verification scores for probabilistic and point forecasts and the coverage and average width of  $77.78\,\%$ central prediction intervals are reported in Table \ref{tab:tab3}. Compared with the raw ensemble and climatology post-processed forecast exhibit the same behavior as before: improved predictive skills and better calibration. In general, models based on combinations of both investigated distributions outperform the TN and LN methods, the smallest CRPS and MAE values and the best coverage, combined with a rather narrow central prediction interval, belong to the regime-switching approach, while the mixture model with parameters optimizing the mean CRPS provides the lowest twCRPS and RMSE scores.

\begin{table}
 \begin{center}
  \caption{Values of the test statistics $t_N$ of the two-sided DM test of equal predictive performance \eqref{eq:DM} for the comparison of the TN-LN mixture models and the benchmark models for the UWME data. Negative values indicate a superior predictive performance of the mixture model in the left column, and positive values indicate a better performance of the benchmark model. 
  Values that are significant at the 0.05 level under the null hypothesis of equal predictive performance are printed in bold.} \label{tab:DM-UWME}
  
  \vskip .5 truecm
  \begin{tabular}{lccc}
    & TN & LN & TN-LN r.s. \\ \hline 
   \multicolumn{4}{l}{\textit{DM test based on LogS}} \\ 
   \hline
   TN-LN mix. (CRPS) & \textbf{-7.63} & \textbf{-13.62} & \textbf{-4.68} \\
   TN-LN mix. (ML) & \textbf{-18.44} & \textbf{-16.45} & \textbf{-12.24} \\
   \hline
     \multicolumn{4}{l}{\textit{DM test based on CRPS}} \\ 
   \hline
   TN-LN mix. (CRPS) & \textbf{-15.73} & \textbf{-16.70} & -0.95 \\
   TN-LN mix. (ML) & \textbf{-5.62} & \textbf{-6.52} & \textbf{2.24}\\
   \hline
     \multicolumn{4}{l}{\textit{DM test based on twCRPS with threshold $r = 9$}} \\ 
   \hline
   TN-LN mix. (CRPS) & \textbf{-7.68} & \textbf{-4.63} & 1.03 \\
   TN-LN mix. (ML) & \textbf{-7.04} & \textbf{-5.11} & 1.06 \\
  \end{tabular}
 \end{center}
\end{table}

The results of DM tests of equal predictive performance are summarized in Table \ref{tab:DM-UWME}. In terms of all employed scoring rules, the mixture models exhibit significantly better predictive performance compared to the TN and LN models. As before, the results for the comparisons with the TN-LN combination model are mixed and depend on the scoring rule. While the mixture models show significantly better results in terms of the LogS, the combination model is preferred in terms of the CPRS and its threshold-weighted version.

\begin{figure}[t!]
\begin{center}
\leavevmode
\epsfig{file=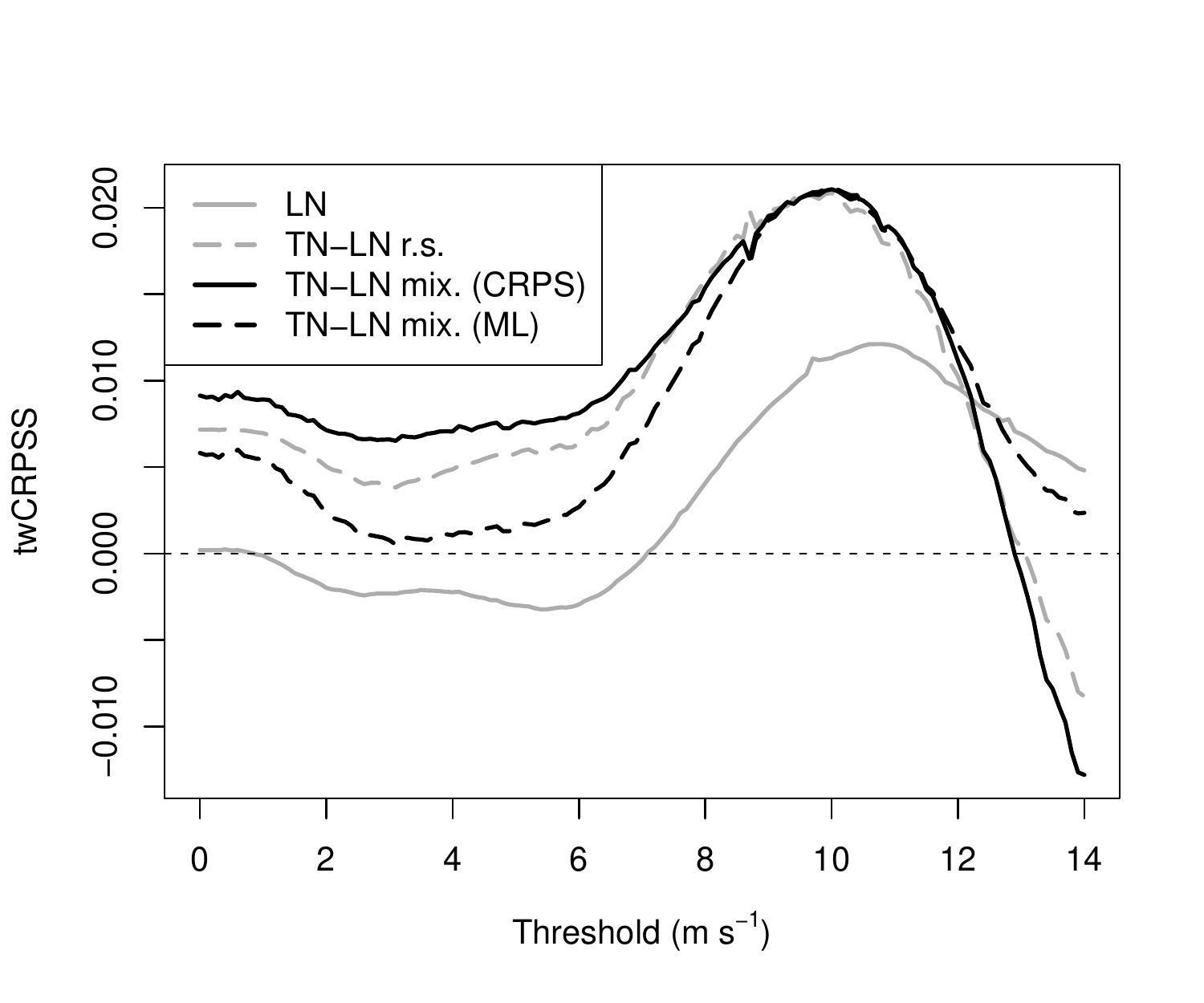,height=7.3cm, angle=0}
\caption{twCRPSS values for the UWME with TN as reference model.}   
\label{fig:fig9}
\end{center}
\end{figure} 

This desirable behavior of the mixture models for high  wind speeds can also be observed in Figure \ref{fig:fig9} where the twCRPSS values of the LN, TN-LN regime-switching and mixture models with respect to the TN EMOS reference model are plotted as functions of the threshold. Up to threshold $r=9$ m\,s$^{-1}$ the regime-switching method slight\-ly outperforms the mixture model, whereas above it this advantage disappears.

\begin{figure}[t!]
\begin{center}
\leavevmode
\epsfig{file=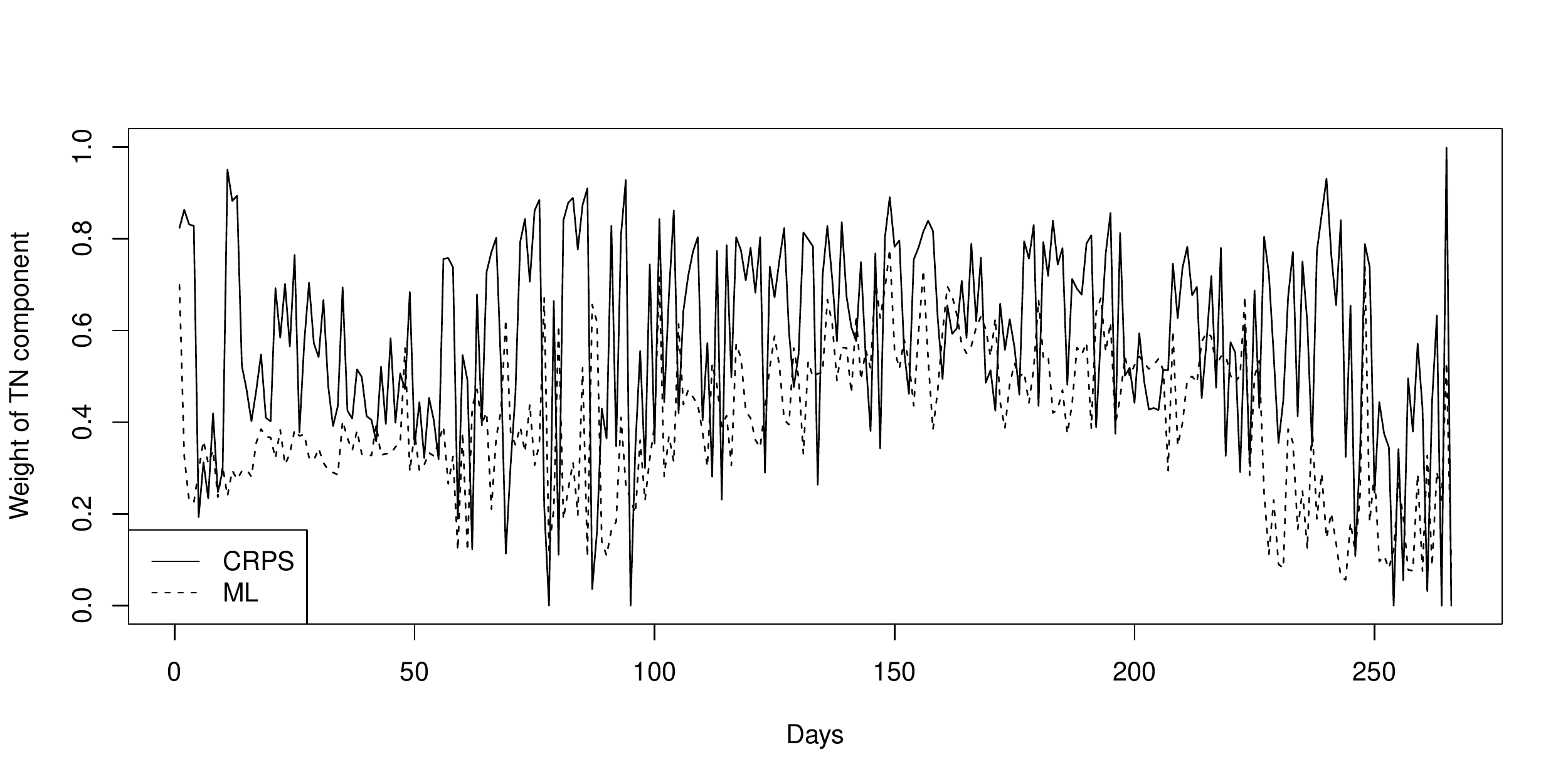,height=7.5cm, angle=0}
\caption{Weights of the TN component for the UWME.}   
\label{fig:fig10}
\end{center}
\end{figure}

In contrast to the ECMWF and ALADIN-HUNEPS ensembles, the weights of the TN component of the two versions of model \eqref{eq:eq3.6} plotted in Figure \ref{fig:fig10} show a positive correlation of $0.214$. Finally, for the UWME the parameter estimates of \ $\mu_{LN}$ \ and \ $\sigma_{LN}$ \ exhibit strong\-er correlations than the estimated location and scale parameters  \ $\mu_{TN}$ \ and \ $\sigma_{TN}$ \ of the TN component, the corresponding values are $0.858, \ 0.826$ and $0.427, \ 0.259$, respectively.

\section{Conclusions}
  \label{sec:sec5}

A new EMOS model for post-processing ensemble forecasts of wind speed is introduced, where the predictive PDF is a weighted mixture of a truncated normal and a log-normal distribution with location and scale/shape parameters depending on the ensemble. Model parameters and mixture weight are estimated simultaneously by optimizing either the mean continuous ranked probabilistic score or the mean logarithmic score (ML estimation) of the predictive distribution over the training data.

The mixture models are tested on three data sets of wind speed forecasts which differ in the generation of the ensemble members and the predicted wind quantities. The predictive skills of the new model are compared with those of the TN based EMOS method \citep{tg}, the LN and the TN-LN regime-switching EMOS models \citep{bl}, the raw ensemble and the climatological forecasts with the help of graphical tools, appropriate verification scores and formal statistical tests of calibration and equal predictive performance. The presented case studies clearly show that compared with the raw ensemble and climatology, statistical post-processing results in a significant improvement in calibration of probabilistic and accuracy of point forecasts. 

In a comparative view of the different EMOS models it can be observed that the TN-LN regime-switching combination model and the new mixture models significantly outperform the simple EMOS models based on single TN and LN distributions and provide much better calibrated probabilistic forecasts as demonstrated by formal statistical tests. The novel mixture models are able to keep up with the TN-LN regime-switching combination method in terms of the various verification scores. Further, the results of formal tests of uniformity indicate a superior calibration of the forecasts produced by the mixture models compared with the combination model. No substantial difference can be observed between the results corresponding to the two parameter estimation methods of the mixture model, ML estimation results in slightly worse verification scores but provides better calibrated forecasts.

Compared with the TN-LN regime-switching combination model, the proposed mixture models exhibit desirable properties from both a theoretical as well as an applied perspective. They are more flexible in that they do not require the exclusive choice of one of the parametric families as forecast distribution. Further, it is not necessary to determine suitable covariates for the model selection, or to estimate the model selection threshold over a training period.

\bigskip
\noindent
{\bf Acknowledgments}

Essential part of this work was made during a visit of S\'andor Baran
at the Heidelberg Institute for Theoretical Studies. The research stay in
Heidelberg was funded by the DAAD program ``Research Stays for University
Academics and Scientists, 2015''.
S\'andor Baran was also supported by
the J\'anos Bolyai Research Scholarship of the Hungarian Academy of
Sciences.
Sebastian Lerch gratefully acknowledges support by the Volkswagen
Foundation within the program
``Mesoscale Weather Extremes -- Theory, Spatial Modelling and
Prediction (WEX-MOP)'', and by the Klaus Tschira Foundation.
The authors thank Tilmann Gneiting, Alexander Jordan and Fabian Kr\"uger
for helpful discussions, and Fabian Kr\"uger for providing \texttt{R}
code for the $\alpha_{1234}^0$ test.
The authors further thank the University of Washington MURI group for
providing the UWME data and Mih\'aly Sz\H ucs from the HMS for providing
the ALADIN-HUNEPS data.

\end{document}